\documentclass[prl,aps,twocolumn,showpacs]{revtex4}%
\usepackage{amsmath}
\usepackage{amsfonts}
\usepackage{amssymb}
\usepackage{graphicx}%
\providecommand{\U}[1]{\protect\rule{.1in}{.1in}}
\begin{document}
\title{Intrinsic magnetic properties of NdFeAsO$_{0.9}$F$_{0.1}$ superconductor from local and global measurements}
\author{R.~Prozorov}
\email{prozorov@ameslab.gov}
\affiliation{Ames Laboratory and Department of Physics \& Astronomy, Iowa State University,
Ames, IA 50011}
\author{M.~E.~Tillman}
\affiliation{Ames Laboratory and Department of Physics \& Astronomy, Iowa State University,
Ames, IA 50011}
\author{E.~D.~Mun}
\affiliation{Ames Laboratory and Department of Physics \& Astronomy, Iowa State University,
Ames, IA 50011}
\author{P.~C.~Canfield}
\affiliation{Ames Laboratory and Department of Physics \& Astronomy, Iowa State University,
Ames, IA 50011}
\date{13 May 2008}

\begin{abstract}
Magneto-optical imaging was used to study the local magnetization in
polycrystalline NdFeAsO$_{0.9}$F$_{0.1}$ (NFAOF). Individual crystallites up to
$\sim200\times100\times30$ $\mu m^{3}$ in size could be mapped at various
temperatures. The in-grain, persistent current density is about $j\sim10^{5}$
A/cm$^{2}$ and the magnetic relaxation rate in a remanent state peaks at about
$T_{m}\sim38$ K. By comparison with with the total magnetization measured in a
bar-shaped, dense, polycrystalline sample, we suggest that NdFeAsO$_{0.9}$F$_{0.1}$ is similar to a layered high-$T_{c}$, compound such as
Bi$_{2}$Sr$_{2}$CaCu$_{2}$O$_{8+x}$ and exhibits a $3D\rightarrow2D$ crossover
in the vortex structure. The $2D$ Ginzburg parameter is about $Gi^{2D}%
\simeq10^{-2}$ implying electromagnetic anisotropy as large as $\epsilon
\sim1/30$. Below $T_{m}$, the static and dynamic behaviors are consistent with
collective pinning and creep.

\end{abstract}

\maketitle

The recently discovered rare earth iron oxypnictides represent a new class of
layered, high-$T_{c}$ superconductors \cite{Kamihara2008}. Rich chemistry
allows for both electron and hole doping \cite{wen2008} as well as magnetism
coexistingwith, and possibly enhancing superconductivity. Unfortunately, the
same rich chemistry makes growth difficult and, so far, only polycrystalline samples exist with the exception of a very recent report of Sm - based single crystals of sizes smaller than reported in this work \cite{zhigadlo2008}.
NdFeAsO$_{0.9}$F$_{0.1}$ (NFAOF) with transition temperature exceeding $51$
K was reported in Ref.\cite{ren-2008}. This and related materials with Sm and
Pr have the highest ambient pressure $T_{c}$ values and are also interesting
because of the possible interplay of the local moment of the rare earth and
superconductivity. Predictions have been made for vortex melting similar to
Bi$_{2}$Sr$_{2}$CaCu$_{2}$O$_{8+x}$ (BSCCO-2212) \cite{lv-2008} and quantum
critical behavior \cite{giovannetti-2008}. A recent NMR study favors nodal
superconductivity \cite{Grafe2008}, infrared ellipsometry suggests large
electromagnetic anisotropy \cite{Dubroka2008} and evidence of electromagnetic
granularity has also been reported \cite{Yamamoto2008}. These observations
prompt further investigation of similarities to high-T$_{c}$ cuprates. So far,
reported magnetic measurements are limited to indication of the transition
temperature. In this Letter we present detailed magnetization measurements on both single crystallite and bulk, polycrystalline NFAOF. These results provide clear insight into the mesoscopic and macroscopic behavior of the superconducting mixed state.  NFAOF exhibits fast non-monotonic magnetic relaxation of the vortex state that is strikingly similar to that associated with the $2D$ vortex melting found in BSCCO-2212, but NFAOF has much stronger interlayer coupling.

High pressure synthesis of samples with a nominal composition Nd(O$_{0.9}%
$F$_{0.1}$)FeAs was carried out in a cubic, multianvil apparatus, with an edge
length of $19$ mm from Rockland Research Corporation. Stoichiometric amounts
of NdFe$_{3}$As$_{3}$, Nd$_{2}$O$_{3}$, NdF$_{3}$ and Nd were pressed into a
pellet with mass of approximately $0.5$ g and placed inside of a BN crucible
with an inner diameter of $5.5$ mm. The synthesis was carried out at about
$3.3$ GPa. The temperature was increased over a period of one hour to
$1350-1400$ $^{o}$C and held for $8$ hours before being quenched to room
temperature. The pressure was then released and the sample removed
mechanically. More details of the synthesis and characterization will be found
elsewhere \cite{Tillman2008}. The value of 10\% F substitution is nominal,
based on the initial stoichiometry of the pellet. The synthesis yields
polycrystalline NdFeAsO$_{0.9}$F$_{0.1}$ samples that contain what appears
to be plate-like single crystals as large as 300 $\mu m$. Whereas extraction
of these crystallites is difficult, we could measure properties of individual
crystals by using local magneto-optical imaging. Comparing our observation
with data from conventional magnetometery, we are able to provide in-depth
magnetic characterization of these new superconductors. For the measurements
of total magnetic moment, a slab-like sample that showed best overall
screening and trapping of the magnetic flux was selected by using
magneto-optical imaging. \textit{Quantum Design} MPMS magnetometer was used
for the measurements of the total magnetic moment. Magneto-optical (MO)
imaging was performed in a $^{4}$He optical flow-type cryostat using Faraday
rotation of polarized light in a Bi - doped iron-garnet film with in-plane
magnetization \cite{prozorov2007d}. The spatial resolution of the technique is
about 3 $\mu m$ with a sensitivity to magnetic field of about 1 G. The
temporal resolution is about 30 msec.%

\begin{figure}[ptb]
\begin{center}
\includegraphics[
height=7.1873cm,
width=8.5712cm
]%
{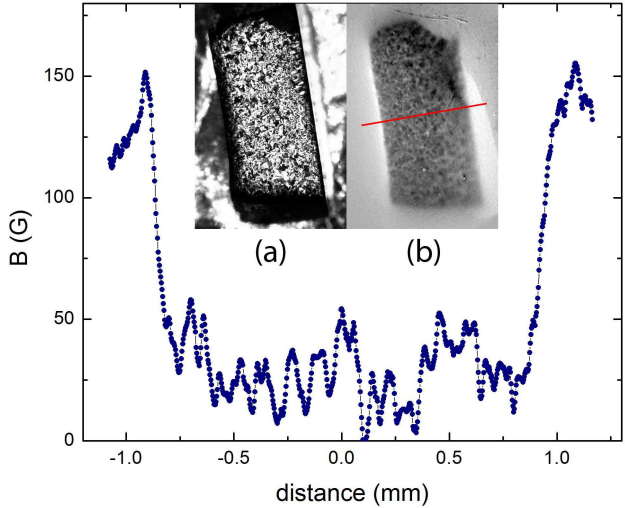}%
\caption{Main frame: Meissner screening of a 100 Oe field applied after zfc.
$B(r)$ was taken along the line shown in inset (b). Inset (a): Optical image
of the sample used in magnetometry and local studies. Inset (b): MO\ image of
a 100 Oe field applied after zfc. }%
\label{fig1}%
\end{center}
\end{figure}

\begin{figure}
[ptbh]
\begin{center}
\includegraphics[
height=12.1737cm,
width=8.1012cm
]%
{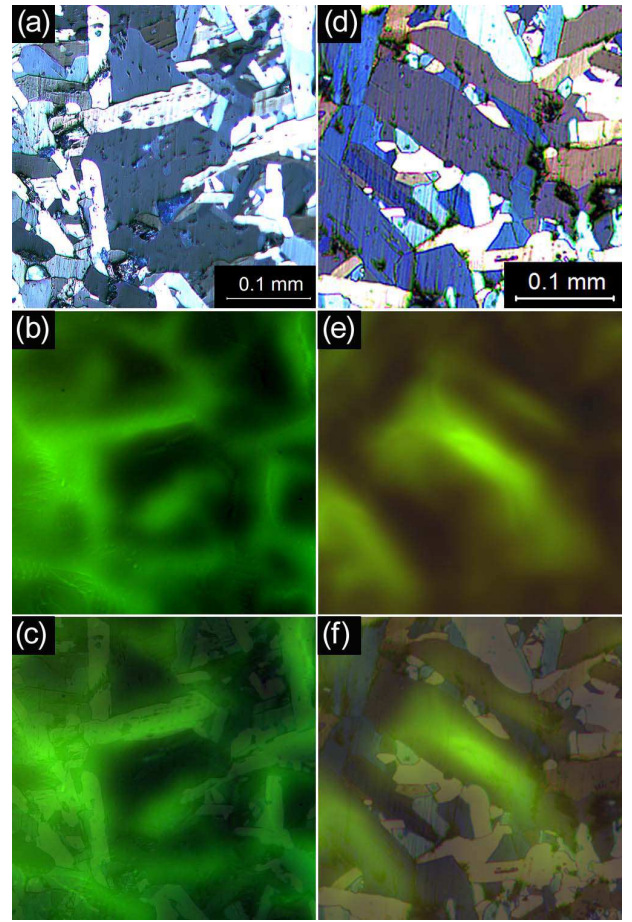}%
\caption{(a) polarized-light image of part of the polished sample surface. (b)
distribution of the magnetic induction upon penetration into the
superconducting state, 550 Oe and 5 K is shown. (c) Superposition of (a) and
(b). (d) - different part of polished surface. (e) remanent (trapped) flux
after fc and turning field off. (f) Superposition of (d) and (e).}%
\label{fig2}%
\end{center}
\end{figure}

We first examined morphological features visible in polarized-light and then
establish the correspondence between these features and the observed local
magnetic behavior. Figure \ref{fig1} (a) shows the entire sample used in the
comparative study. Figure \ref{fig1} (b) shows the MO\ image in a $100$ Oe
applied magnetic field obtained after zero-field cooling (zfc). There is
Meissner screening, better seen in the magnetic induction profile shown in the
main frame of Fig. \ref{fig1} taken along the line shown in Fig. \ref{fig1}
(b). Figures \ref{fig2} (a) and (d) show two different regions of the polished
sample and provide clear evidence of well - faceted crystallites with
cross-sections as large as $200\times100$ $\mu m^{2}$. Sensitivity to the
orientation of the light polarization plane with respect to the crystal
structure serves as an additional indication that we are dealing with well
ordered crystallites. Tetragonal symmetry of the unit cell suggests that the
crystals should grow as plates with the $ab-$ plane being the extended surface
and the $c-$ axis along the shortest dimension. We examined various
cross-sections of the $\simeq5$ mm diameter, $5$ mm height pellets. Based on
the thickness of extremely rectangular grains we estimate the thickness of the
plates to be as large as $30$ $\mu m$. The magnetic flux penetration is
consistent with isotropic in-plane persistent current densities, further
confirming that the large grains represent the $ab-$ plane of the tetragonal crystallites.%

We can correlate this microstructure with the ability of this superconductor
to shield magnetic field and trap the flux. Figure \ref{fig2} (b) shows
penetration of the magnetic flux into the region imaged in Fig. \ref{fig2} (a)
after zfc, whereas Fig. \ref{fig2} (e) shows remanent (trapped) flux in the
region imaged in Fig. \ref{fig2} (d). Figures \ref{fig2} (c) and (e) are the
superpositions of Figs. \ref{fig2} (a) and (b) and Figs. \ref{fig2} (d) and
(e), respectively. The correspondence between good superconducting regions and
the largest crystals is evident.

\begin{figure}
[ptb]
\begin{center}
\includegraphics[
height=6.3021cm,
width=8.1012cm
]%
{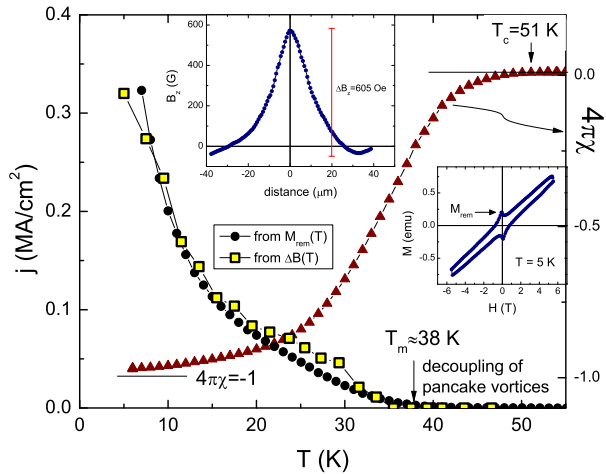}%
\caption{Persistent current density, $j$, estimated from local
MO\ measurements (see text). For comparison, $M_{rem}\left(  T\right)  $ is
scaled to match the amplitude (circles). Also shown is $4\pi\chi$ (triangles,
right axis) to emphasize the "magnetic" $T_{c}$. Upper inset shows example
$B\left(  r\right)  $ and the definition of $\Delta B$; Lower inset shows an
$M\left(  H\right)  $ loop at $5$ K.}%
\label{fig3}%
\end{center}
\end{figure}

There are several ways to estimate the shielding current from the measurements
of total magnetic moment. For a slab of dimensions $2w\times2b\times2d$
(magnetic field is along the $d$ side and $w\leq b$) within the Bean model
\cite{BEAN1964}, total magnetic moment, $M$, is given by $M=jwV\left[
1-w/\left(  3b\right)  \right]  /2c$ where, $c$ is the speed of light, $V$ is
sample volume and $j$ is the persistent (Bean) current that results from
vortex pinning (we avoid calling this quantity "critical" current, because it
is significantly affected by the magnetic relaxation). In the present case,
for magnetic field parallel to the long side of a slab, $2d=0.44$ cm,
$2w=0.07$ cm and $2b=0.18$ cm and $V=5.54\times10^{-3}$ cm$^{3}$. Therefore,
$j\left[  \text{A cm}^{-2}\right]  \simeq1.2\times10^{5}\times M\left[
\text{emu}\right]  $. Taking the maximum half-width of the full hysteresis
loop, see inset to Fig.\ref{fig3}, at $H=0$, $M_{rem}\left(  5\ \text{K}%
\right)  \simeq0.2$ emu, we estimate $j\left(  5\ \text{K}\right)
\simeq2.4\times10^{4}\ $A cm$^{-2}.$ Another way to estimate shielding current
density is to measure the field of full penetration \cite{ProzorovPhD2008},
$H^{\ast}={4wj}\left(  {2\arctan\left(  \eta\right)  +\eta\ln\left(
{1+\eta^{-2}}\right)  }\right)  /c,$ where $\eta=d/w.$ Field of full
penetration may be estimated from the minimum of the $M\left(  H\right)  $
loop. In our case, we have $\eta=6.3$, $H^{\ast}\simeq1100$ Oe and therefore,
${j\simeq}2.6\times10^{4}$ A cm$^{-2}$, which is close to the above estimate
and implies that superconducting fraction is close to $100$ \%. To verify this
conclusion we measured reversible (Meissner) $M\left(  H\right)  $ at small,
$+/-$ 10 Oe, field span which showed no hysteresis. The overall shielding was
then estimated from $4\pi\chi=4\pi\left(  1-N\right)  M/VH$, where $N=\left(
1+2\eta\right)  ^{-1}\simeq7.4\times10^{-2}$ is the demagnetization factor (we
neglected London penetration depth, $\lambda$, that enters via $\lambda/w$
correction and thus irrelevant \cite{Prozorov2000b}). At $5$ K we found,
$4\pi\chi=-0.98$, essentially perfect diamagnetism. Given the uncertainty in
$V$ and $N$, as well as the neglected $\lambda$ we see that this sample
exhibits close to $100$ \% diamagnetic screening.%

These results can be compared to local magnetic measurements. The persistent
current may be evaluated from the measurements of the $z-$ component of the
magnetic induction on the slab's top surface using magneto-optical imaging.
Defining $\Delta B=\left\vert B_{z}\left(  x=0,z=d\right)  -B_{z}\left(
x=w,z=d\right)  \right\vert $ and assuming full critical state, we have
\cite{ProzorovPhD2008}%

\[
\frac{\Delta Bc}{4\widetilde{{j}}\widetilde{{w}}}={\eta\ln\frac{{\left(
{1+4\eta^{2}}\right)  ^{2}}}{16\eta^{3}\sqrt{1+\eta^{2}}}+2\arctan\left(
{2\eta}\right)  -\arctan\left(  \eta\right)  }%
\]
where we use "$\sim"$ to refer to quantities estimated for the individual
crystallites. In the present case, a rectangular crystallite of $2\widetilde{w}=70$ $\mu$m, $2\widetilde{b}=160$ $\mu$m and assuming thickness, $2\widetilde {d}=30$ $\mu$m, was
analyzed. Therefore we have ${\eta\simeq30/70\simeq0.43}$ and $\widetilde{{j}%
}\left[  \text{A cm}^{-2}\right]  {\simeq529}\Delta B\left[  \text{G}\right]
$. With $\Delta B\simeq605$ we estimate $\widetilde{j}\left(  5\ \text{K}%
\right)  \simeq3.2\times10^{5}$ A cm$^{-2}$. Clearly, there is about an order
of magnitude difference with the estimate from bulk magnetization. The reason
is simple, - critical state is established in each individual crystallite, not
on the scale of the entire sample, so and there is no macroscopic Bean
gradient of the magnetic induction. This is why local measurement on the scale
of an individual grain are required. Note that if $2\widetilde{d}$ were
infinite, the conversion would be $\widetilde{{j}}{\simeq455}\Delta B$, and
for smaller $2\widetilde{d}$ we expect larger conversion factor. Therefore, we
provided the conservative estimate of the critical current.

Figure \ref{fig3} summarizes \emph{local} and global measurements of the
persistent current density in NFAOF. Squares show $\widetilde{j}\left(
T\right)  $ obtained from $\Delta B$ as defined in the upper inset. For
comparison, full circles show the temperature dependence of the remanent
magnetization rescaled by a single scaling factor to match the $\widetilde
{j}\left(  T\right)  $ obtained from local measurements. Suppose that the slab
was split into crystallites of width $2\widetilde{w}$ each carrying
$\widetilde{j}\sim10^{5}$ A cm$^{-2}$. A total of $n=w/\widetilde{w}$
crystallites would produce a total magnetic moment of $\widetilde{M}\sim
n\widetilde{j}\widetilde{w}\widetilde{V}\sim\widetilde{j}\widetilde{w}$. If we
want this moment to match the observed $M$, then $\widetilde{w}\sim
wj/\widetilde{j}\sim0.1w$. With $w\simeq350$ $\mu m$ we obtain a good
agreement with the directly observed width of the crystallites, $\widetilde
{w}\sim35$ $\mu m$. This yields an important conclusion - in our sample global
magnetic measurements can be used to access intra-grain persistent current,
but the estimated magnitudes will be about $10$ times lower.

Figure \ref{fig3} also shows other important features. The triangles (right
axes) show magnetic susceptibility, $4\pi\chi$, measured at $H=10$ Oe. The
magnetic $T_{c}\simeq51$ K and resistive $T_{c}\simeq53$ K \cite{Tillman2008}.
However, the $\widetilde{j}\left(  T\right)  $ virtually vanishes above about
$35$ K. Remarkably, the order of magnitude and overall temperature dependence
of $\widetilde{j}\left(  T\right)  $ is quite similar to that observed in
BSCCO-2212 \cite{Prozorov2003} where there is a clear crossover above $\sim30$
K associated with decoupling of pancake vortices ($3D\rightarrow2D$ crossover)
\cite{Pradhan1994}.

\begin{figure}
[ptbh]
\begin{center}
\includegraphics[
height=6.4229cm,
width=8.1012cm
]%
{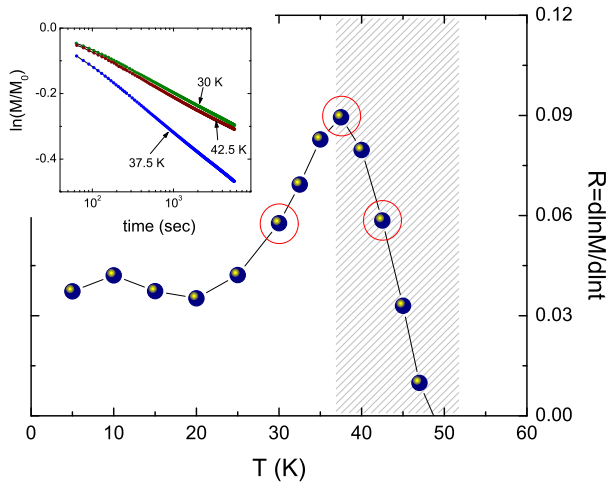}%
\caption{Logarithmic relaxation rate $R=d\ln M/d\ln t$ as function of
temperature. Shadowed region shows anomalous behavior. Inset: time - dependent magnetization at three (circled) temperatures below, at and above the peak in $R\left(  T\right)  $. }%
\label{fig4}%
\end{center}
\end{figure}

Given this last observation, we now turn to a discussion of the sample's
dynamic properties. Magnetic relaxation is a valuable tool for determining
vortex-related parameters of a superconductor \cite{Blatter1994,Yeshurun1996}.
The relaxation rate depends on the pinning parameters as well as structure of
the Abrikosov vortices and vortex lattice. A perfunctory inspection of the
magnetic relaxation reveals a very large time dependence even at low
temperatures. At 5 K, there is a $16\%$ change of total magnetic moment over
1.5 h; there is even larger change of $38\%$ at $\sim38$ K. Whereas one can
use sophisticated, nonlinear models, here it is sufficient to examine the
logarithmic relaxation rate, $R=\left\vert d\ln M/d\ln t\right\vert $ which
allows for comparison with other systems and is sample volume independent.
Figure \ref{fig4} shows a sharp increase of $R\left(  T\right)  $ and a peak
at about $\sim38$ K. It is worth noting that temperature of the peak is in the
same temperature range that the apparent persistent current (Fig.\ref{fig3})
drops almost to zero. Both observations are consistent with the vanishing of
the barrier for vortex escape from the pinning potential at this temperature.%

Comparing to cuprate, high-$T_{c}$ superconductors, we note that for different
YBCO samples, including flux-grown, melt-processed crystals and films,
$R\left(  T\right)  $ is confined to a narrow range between $0.02$ and $0.04$
and is fairly flat \cite{Yeshurun1996}. In contrast, for BSCCO, $R\left(
T\right)  $ exhibits a peak at temperature of a crossover associated with
pancake decoupling \cite{Pradhan1994,Yeshurun1996}. Examining the magnetic
field dependence of the relaxation rate we find that below $38$ K it drops
very fast, despite the fact that the apparent irreversible magnetization is
almost field independent above 1 kOe. This is consistent with the collective
pinning and creep model in which the barrier for magnetic relaxation rapidly
decreases with increasing magnetic field due to growth of vortex bundles.%

\begin{figure}
[ptbh]
\begin{center}
\includegraphics[
height=6.39cm,
width=8.1012cm
]%
{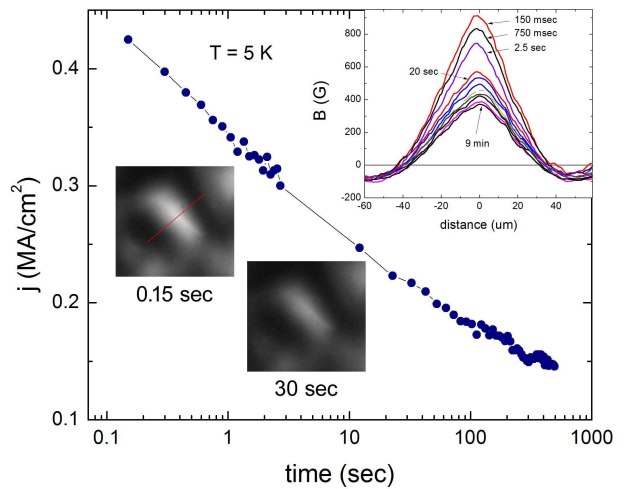}%
\caption{Relaxation of the persistent current measured from the decay of the
magnetic induction profiles shown in the upper inset. Embedded images show two
frames at 0.15 sec and 30 sec after magnetic field was turned off. A line
shows where profiles were taken.}%
\label{fig5}%
\end{center}
\end{figure}

The relaxation can also be measured in the individual crystallites. Figure
\ref{fig5} shows magnetic relaxation measured from a series of the snapshots
taken initially at $150$ msec intervals and later at $10$ sec intervals.
Examples of two such snapshots are shown as insets. To estimate the persistent
current, profiles of magnetic induction, shown in the upper inset, were
measured along the indicated path and then converted into $j$ as described
above. The relaxation rate is quite high, approaching $R=0.1$, which reflects
the fact that effective time constant depends on the sample size.

To understand our results from the point of view of unconventional, layered
superconductors, we note that the overall, observed, magnetic behavior is
similar to anisotropic high-$T_{c}$ cuprates. Although without separate single
crystals it is difficult to make firm conclusions, if we adopt the concept of
a layered superconductor, then the dimensional crossover temperature signifies
$2D$ melting, which, at zero field, is given by $T_{m}^{2D}=\left(
T_{BKT}/4\sqrt{3}\pi\right)  T_{c}/\left(  T_{c}-T_{BKT}\right)  $ where
$T_{BKT}$ is Berezinskii-Kosterlitz-Thouless temperature \cite{Blatter1994}.
In BSCCO-2212, $T_{BKT}\simeq82$ K, $T_{c}\simeq95$ K and $T_{m}^{2D}\simeq28$
K. For NFAOF we therefore obtain, $T_{BKT}\simeq48$ K with the observed
$T_{c}\simeq51$ K and $T_{m}^{2D}\simeq38$ K. This lets us estimate the
Ginzburg parameter, $Gi^{2D}\simeq\left(  1-T_{BKT}/T_{c}\right)
v_{s}^{\infty}/2\sqrt{2}\simeq10^{-2}$ which is consistent with our assumption
of a layered superconductivity ($v_{s}^{\infty}\simeq0.5-0.9$ is the coupling
renormalization \cite{Blatter1994}). This result implies a rather small
penetration depth, $\lambda\left(  0\right)  <100$ nm and electromagnetic
anisotropy of the order of $\epsilon\sim1/30$. These estimates will have to be
refined when separate, single crystals are available.

In conclusion, we found evidence for a sharp crossover in the pinning
mechanism at around $T_{m}^{2D}\approx38$ K, which we interpret as a crossover
from $3D$ Josephson coupled pancake vortices to decoupled 2D pancakes, similar
to that found at $\sim30$ K in BSCCO \cite{Pradhan1994}. The temperature and
field dependence of the persistent current as well as its logarithmic
relaxation rate are consistent with collective pinning and creep
\cite{Blatter1994,Yeshurun1996}. The estimated current density, $\sim10^{5}$ A
cm$^{-2}$ is low enough for weak collective pinning to hold. Overall, the
behavior of NFAOF is somewhere in between YBCO and BSCCO.

Discussions with S. Bud'ko, V. Kogan, M. Tanatar, A. Kaminskii, W. McCallumn,
K. Dennis and J. Schmalian are greatly appreciated. Work at the Ames
Laboratory was supported by the Department of Energy-Basic Energy Sciences
under Contract No. DE-AC02-07CH11358. R. P. acknowledges support from NSF
grant number DMR-05-53285 and the Alfred P. Sloan Foundation.

\end{document}